\documentclass[twocolumn,
nofootinbib,
prd,aps,
tightenlines]{revtex4}
\usepackage{color}
\usepackage{bm}
\usepackage[]{youngtab}
\usepackage{pstricks}

\def\nn{ \nonumber \\ }
\def\ket#1{ \left| #1 \right\rangle }
\def\abs#1{ \left| #1 \right| }
\def\lchi{\Lambda_{\chi}}

\begin{document}

\title{The Chiral Soliton Model for Arbitrary Colors and Flavors}

\author{Elizabeth~Jenkins}
\author{Aneesh V.~Manohar}
\affiliation{Department of Physics, University of California at San Diego,
  La Jolla, CA 92093\vspace{4pt} }

\date{\today}
\begin{abstract}
The quantum numbers of the chiral soliton are derived for an arbitrary number of colors and flavors.
\end{abstract}
\maketitle

The discovery of the $\Theta^+$ baryon~\cite{Thetadiscovery}  has led to renewed interest in the chiral soliton model for baryons. The connection between the chiral soliton and quark models is clear if they are studied as a function of the number of colors and flavors.  In this paper, we derive the states obtained by collective coordinate quantization of the chiral soliton model~\cite{Skyrme}, for an arbitrary number of colors $N_c$ and flavors $F$. 

QCD has a $SU(F)_L \times SU(F)_R$ chiral symmetry, which is spontaneously broken to the diagonal $SU(F)$ flavor group. The Goldstone bosons are elements of an $SU(F)$ matrix $U(\mathbf{x},t)$, and the dynamics is given by a 
chiral Lagrangian $L_\chi$. The topology of the $SU(F)$ manifold allows for the possibility of solitons. The standard hedgehog configuration for this (static) soliton is
\begin{eqnarray}
U_0(\mathbf{x}) &=& \left( \begin{array}{c|c} e^{i \mathbf{\tau \cdot \hat x} F(r)} & 0 \\
\hline
0 & \openone \end{array} \right),
\label{1}
\end{eqnarray}
with non-trivial fields only in the upper $2 \times 2$ block of the $F \times F$ matrix $U$.
The shape function $F(r)$ is determined by solving the non-linear classical field equations of $L_\chi$. The soliton has winding number one, and has been argued to have baryon number one~\cite{Witten}, even though it is made up purely of meson fields.

The chiral Lagrangian has an expansion in powers of $\partial/\lchi$, where $\lchi\sim 4 \pi f_\pi$ is the scale of chiral symmetry breaking, so $F(r)$ varies over a typical scale 
$r \sim \lchi^{-1}$.  All space derivative terms in $L_\chi$ are equally important,  and one cannot determine the shape (or even whether the soliton is stable) from the first few terms in $L_\chi$. Nevertheless, assuming the existence of the soliton  with some arbitrary shape function $F(r)$ allows one to compute the quantum numbers of the low-lying states in the  baryon spectrum in the $N_c \to \infty$ limit. Since the mass and moment of inertia of the soliton are both of order $N_c$, the low-lying states are given by quantizing the rotational motion of the soliton, and have mass-splittings relative to the lowest baryon state of order $1/N_c$. The semiclassical expansion of the effective theory is an expansion in powers of $1/N_c$, or equivalently, in powers of time-derivatives.

The collective coordinates for the standard soliton configuration are translations, space rotations and flavor rotations. Translations produce a $\mathbf{P}^2/2M$ shift in the energy, but do not affect the quantum numbers of the soliton, and will be neglected here. Space rotations are generated by $J^i$, and flavor rotations by $T^a$. The flavor generators are normalized to $\text{Tr}\, T^a T^b = \delta^{ab}/2$ in the fundamental representation.
We will need the decomposition of the flavor group $SU(F) \to SU(2) \times SU(F-2) \times U(1)$, where
$SU(2)$ isospin acts on the first two flavors and is generated by $I^a$; $SU(F-2)$ acts on the remaining flavors and is generated by $S^a$; and the $U(1)$ generator is $T_Y =  \sqrt{(F-2)/(4F)} \mathcal{Y}$,
\begin{eqnarray}
\mathcal{Y} &=& \text{diag}(1,1,y,\ldots,y), \qquad y = -{2 \over F-2}.
\label{2}
\end{eqnarray}
For three flavors, $SU(F-2)$ is absent, and $\mathcal{Y}=3Y$, where $Y$ is the usual $SU(3)$ hypercharge. For two flavors, $SU(F-2)$ and $U(1)$ are both absent.

The soliton in a rotated configuration is described by the matrix $A \in SU(F)$, where $U =  A U_0(\mathbf{x})A^{-1}$. Transforming the soliton  by the flavor transformation $V \in SU(F)$ gives $A \to V A$, and transforming by the space rotation $W$ gives $A \to A W^{-1}$. Space rotations are equivalent to right-multiplication of $A$ by $W^{-1}$, 
 because spin and isospin are locked together by the $\bm{\tau} \cdot \mathbf{x}$ dependence, and the soliton Eq.~(\ref{1}) satisfies $(\mathbf{I}+\mathbf{J})U_0(\mathbf{x})=0$.

Quantizing the collective coordinate $A$ leads to a tower of states. Different collective coordinates $A$ and $Ah$ lead to the same soliton configuration $U$ if $U_0 = h U_0 h^{-1}$. The elements $h$ which leave the soliton invariant form the little-group of the solution, and are given by $\mathbf{I+J}$, $S^a$, and $\mathcal{Y}$. The allowed states of the Skyrme model have wavefunctions  $\sqrt{\dim R}\, D^{(R)}_{ab}(A)$, where $R$ is an irreducible $SU(F)$ representation. The soliton transforms like $\ket{Ra}$ under flavor, where $a$ is the particular element of $R$. The little-group constraint for arbitrary $F$ derived in Ref.~\cite{AM} generalizes the hypercharge constraint for three flavors~\cite{Witten}, and is: Decompose the representation $R$ of $SU(F)$ into representations of $SU(2) \times SU(F-2) \times \mathcal{Y}$. The allowed wavefunctions are those for which the state $\ket{Rb}$ is an $SU(F-2)$ singlet and has $\mathcal{Y}=N_c$. The soliton spin is given by the isospin of $\ket{Rb}$.
The consequences of this constraint are worked out for arbitrary flavors, starting with $F \ge 5$ and then restricting to the special cases $F=2,3,4$.

An irreducible representation of  $SU(F)$ is described by the Dynkin weight $(n_1,n_2,\ldots,n_{F-1})$, i.e.\ a Young tableau with $n_1$ columns of one box, $n_2$ columns with two boxes, etc.\ Each box in the Young tableau corresponds to an (upper) index on the $SU(F)$ tensor. Indices  in a given column are totally antisymmetrized. We will refer to a column with $n$ boxes as a $[n]$ column, to emphasize the antisymmetry in the $n$ indices. A particular state is described by choosing values for each index (or box), i.e. deciding whether to set it to $u$, $d$, $s$, etc.\  For example,
\Yvcentermath1
\begin{eqnarray}
\young(11111,22,3,4)
\end{eqnarray}
is the $T^{111[12][1234]}$ element of the $(3,1,0,1)$ representation of $SU(5)$, and is a state with hypercharge $7+2y=17/3$ using Eq.~(\ref{2}).

The hypercharge constraint says that there must be a state with $\mathcal{Y}=N_c$ in $R$. Each index chosen to be $1,2$ (i.e. $u$, $d$) contributes $1$ to $\mathcal{Y}$, and each index chosen to be $3,\ldots F$ contributes $y<0$. Thus, the minimum number of boxes in the Young tableau is equal to $N_c$, and the $N_c$-box states $\ket{Rb}$ have all indices set equal to $1,2$. Since one can antisymmetrize in at most two indices if they are restricted to have at most two values, the allowed  tableaux can only contain $[1]$ and $[2]$ columns.
The allowed weights are $w=(n_1,n_2,0,\ldots,0)$, with $n_1 + 2 n_2=N_c$, and
the $\ket{Rb}$ states are:
\begin{eqnarray}
&& \young(1111111,222222)  \nn
&& \young(11111111,22222)  \label{3} \\
&& \vdots \nn
&& \young(1111111111111)\nonumber
\end{eqnarray}
The $SU(F-2)$ constraint is satisfied automatically, since no index transforms under $SU(F-2)$, so the state is an $SU(F-2)$ singlet. A $[2]$ column with indices $1$ and $2$ is the antisymmetric combination $ud-du$, and has isospin zero, so the above states have isospin $1/2,3/2,\ldots,N_c/2$.
The $SU(2)$ constraint converts this to  spin, so we have a tower of states
$w=(n_1,n_2,0,\ldots,0)$, with $n_1 + 2 n_2=N_c$, and spin $n_1/2$, which is the usual non-exotic tower of Skyrme states.

There are additional states in the Skyrme tower. The hypercharge constraint can be satisfied with more than $N_c$ boxes, by choosing some boxes to be $1,2$, and the rest to be $3,\ldots F-2$, so that $\mathcal{Y}$ for the additional boxes adds to zero. All additional boxes with values  $3,\ldots F-2$ must form an $SU(F-2)$ singlet. The only way to form an $SU(F-2)$ singlet is to completely antisymmetrize $F-2$ boxes using the $\epsilon$-symbol of $SU(F-2)$, i.e. they must have the form
\Yvcentermath1
\begin{eqnarray}
\young(3,\cdot,\cdot,F) 
\label{4}
\end{eqnarray}
Thus, anytime we add an index $\ge 2$, we also must add one each of all the remaining indices $3,\ldots,F-2$ in an $[F-2]$ column, as well as two more boxes with values $1,2$ to satisfy the hypercharge constraint. The possible tableaux depend on how the $1,2$ boxes are assigned. The $1,2$ boxes could go into $[1]$ or $[2]$ columns or some of them could be added to the $[F-2]$ column, to form a $[F-1]$ or $[F]$ column. The only possible columns allowed are $[1]$, $[2]$, $[F-2]$, $[F-1]$ and $[F]$, and the total number of boxes is $N_c + r^\prime F$, with $r^\prime$ being the total number of $[F-2]$, $[F-1]$ and $[F]$ columns. In $SU(F)$, $[F]$ columns are flavor singlet, and can be thrown away. 
Thus, the allowed tableaux contain only $[1]$, $[2]$, $[F-2]$,  and $[F-1]$ columns, and have $N_c+rF$ boxes, where the integer $r \ge 0$ is equal to the number of $[F-2]$ and $[F-1]$ columns.
The allowed $SU(F)$ weights are $w=(n_1,n_2,0\ldots,0,n_{-2},n_{-1})$, where we have relabelled $n_{F-2} \to n_{-2}$ and $n_{F-1} \to n_{-1}$.  The four positive integers $n_i$
satisfy
\begin{eqnarray}
n_{-1} + n_{-2} &=& r \nn
n_1 + 2n_2 +(F-2)n_{-2}+(F-1) n_{-1} &=& N_c + r F \nn
n_1 + 2 n_2 -2 n_{-2} - n_{-1} &=&N_c 
\label{6}
\end{eqnarray}
where the first relation defines $r$ as the number of $[F-1]$ and $[F-2]$ columns, and the second relation sets the total number of boxes equal to $N_c+ rF$.
The third relation is a linear combination of the first two.

Spin is equal to the isospin under the $SU(2)$ subgroup for the $\mathcal{Y}=N_c$ states. A $[F-2]$ column has the form shown in Eq.~(\ref{4}) and is an $SU(2)$ singlet. A $[F-1]$ column has the form Eq.~(\ref{4}) plus one box set equal to $1,2$, and so transforms as isospin 1/2. The $[2]$ columns are isosinglets, and $[1]$ columns have isospin 1/2.  The $n_1$ $[1]$ columns are completely symmetrized, and so have isospin $n_1/2$. Similarly, the $n_{-1}$ $[F-1]$ columns have isospin $n_{-1}/2$. Thus, the allowed isospins for $\ket{Rb}$, which are the allowed spins, are $j = (n_1/2) \otimes (n_{-1}/2)$. Furthermore, each flavor representation occurs at most once in the collective coordinate quantization. To get multiple copies of the same state, such as two $(\mathbf{8},\frac12)$ states requires vibrational excitations of the soliton. 

The above analysis gives the classification of states for $F \ge 5$ presented in Ref.~\cite{JM1}. Taking the last relation in Eq.~(\ref{6}) modulo two shows that the soliton is a fermion or boson depending on whether $N_c$ is odd or even.  The cases $F=4,3,2$ are all special and must be considered individually.

For $F=4$, one cannot distinguish between $[2]$ and $[F-2]$. The above analysis remains valid for the allowed $SU(4)$ weights 
$w=(n_1,n_0,n_{-1})$, where $n_0 \equiv n_2 + n_{-2}$ is the number of $[2]$ columns.  Eq.~(\ref{6}) becomes
\begin{equation}
n_1 + 2n_0 + 3 n_{-1} = N_c + 4 r ,\ \  n_{-1} \le r \le  n_0 + n_{-1},\ 
\label{7}
\end{equation}
with spins  $(n_1/2)\otimes (n_{-1}/2)$. The first relation modulo two again shows that the soliton is a fermion or boson depending on whether $N_c$ is odd or even. The second condition is necessary and sufficient for there to be two positive integers $n_{\pm 2} \ge 0$ which satisfy $r=n_{-1}+n_{-2}$ and $n_0=n_2+n_{-2}$.  As an example of  why the inequality in Eq.~(\ref{7}) is needed, consider $w=(1,0,2)=\mathbf{\overline{36}}$ for $N_c=3$ so that $r=1$. This representation does not contain any $SU(F-2)$ singlets with $\mathcal{Y}=3$ and is not allowed because it violates the inequality of Eq.~(\ref{7}). 

\begin{figure}
\psset{xunit=0.3cm}
\psset{yunit=0.25981cm}
\psset{runit=0.3cm}
\psset{dotsize=0.2cm}
\begin{pspicture}(-8,-5)(8,15)
\rput(0,14){$p+1$}
\rput(0,-4){$q+1$}
\rput(0,18){$Y$}
\rput(11,4){$I_3$}
\psline(-2,12)(2,12)(7,2)(5,-2)(-5,-2)(-7,2)(-2,12)
\psline[linestyle=dashed,linecolor=blue](-9,2)(9,2)
\psline{->}(-10,4)(10,4)
\psline(0,-3)(0,13)
\psline{->}(0,15)(0,17)
\psdots[dotstyle=o](-2,12)(0,12)(2,12)(3,10)(4,8)(5,6)(6,4)(7,2)(6,0)(5,-2)
\psdots[dotstyle=o](5,-2)(3,-2)(1,-2)(-1,-2)(-3,-2)(-5,-2)
\psdots[dotstyle=o](-5,-2)(-6,0)(-7,2)(-6,4)(-5,6)(-4,8)(-3,10)
\psdots*[dotstyle=o,fillcolor=black](-1,10)(1,10)(2,8)(3,6)(4,4)(5,2)
\psdots*[dotstyle=o,fillcolor=black](4,0)(2,0)(0,0)(-2,0)(-4,0)
\psdots*[dotstyle=o,fillcolor=black](-5,2)(-4,4)(-3,6)(-2,8)
\psdots*[dotstyle=triangle,linecolor=red,fillcolor=red](0,8)
\psdots*[dotstyle=triangle,linecolor=red,fillcolor=red](-1,6)(1,6)
\psdots*[dotstyle=triangle,linecolor=red,fillcolor=red](-2,4)(0,4)(2,4)
\psdots*[dotstyle=triangle,linecolor=red,fillcolor=red](-3,2)(-1,2)(1,2)(3,2)
\end{pspicture}
\caption{$SU(3)$ weight diagram for the $(p=2,q=5)$ representation. The upper edge has $p+1$ weights, and the lower edge has $q+1$ weights. The outermost layer (open circle) has multiplicity one, the next layer (solid circle) has multiplicity two, and the multiplicities increase by one until one gets to a triangular layer (red triangle), after which they stay constant. The horizontal blue dashed line is drawn through the corners of the weight diagram.
\label{fig:weight}}
\end{figure}
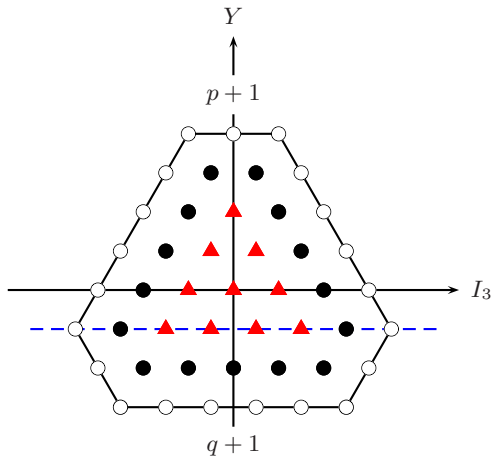

Soliton quantization for $F=3$ has been discussed before~\cite{Witten,Guadagnini,AM,Chemtob,Diakonov}, but the general result derived below is new.
For $F=3$, the general $SU(3)$ representation is  $(p,q)$, which is a traceless tensor $T^{a_1 \ldots a_p}_{b_1 \ldots b_q}$ with $p$ upper and $q$ lower indices. 
The $SU(F-2)$ constraint is absent, and the hypercharge constraint implies that there must be a state with $3Y\equiv \mathcal{Y}=N_c$. The weights of the $(p,q)$ representation in $SU(3)$ have the form shown in Fig.~(\ref{fig:weight}).
The maximum hypercharge is given by choosing all the upper indices of the tensor equal to $1,2$, and the lower indices to $3$, and is given by $3Y_{\rm max}=p+2q$. One moves from a given hypercharge level to the next lower level by replacing one of the upper indices by $3$, or one of the lower indices by $1,2$, so that $3Y \to 3Y-3$. The minimum hypercharge level is reached when the upper indices are all $3$ and the lower indices are all $1,2$, so $3Y_{\rm min}=-2p-q$. The total number of hypercharge levels is $p+q+1$.

The little group constraint is that there must be a $3Y=N_c$ level in the weight diagram, which requires
\begin{eqnarray}
p+2q &=& N_c + 3r, \qquad 0 \le r \le p+q,
\label{8}
\end{eqnarray}
and the states we need are $r$ steps below the maximum $Y$ states. Since $Y_{\text{min}} < 0$ and $N_c >0$, there are always states with $3Y=N_c$ if Eq.~(\ref{8}) is satisfied.  The three integers $p$, $q$ and $r$ are related to the $n_i$ of the general $F \ge 5$ analysis by
\begin{eqnarray}\label{pqr}
p &=& n_1 + n_{-2}, \quad
q = n_{-1} + n_2, \quad
r = n_{-1} + n_{-2} \ .
\end{eqnarray}
It is also convenient to define $s= n_1 + n_2 \equiv p+q -r$.  The minimum spin $j$ is
equal to $j_{\rm min} \equiv \abs{n_1 - n_{-1}}/2 = \abs{p-r}/2 = \abs{s-q}/2$.

To determine the maximum spin, we need to determine the maximum isospin of the states $r$ steps below the maximum $Y$ states. The horizontal dashed line in Fig.~\ref{fig:weight} through the corners of the weight diagram is $q$ steps below the upper edge. If $r \le q$, then the states we need are above (or on) the horizontal dashed line, and if $ r \ge q$, the states are below (or on) the horizontal dashed line, so we consider the two cases separately.  {\it Case ($r \le q$):} The upper edge states have isospin $I=p/2$. As one drops down in $Y$, one gets states with different values of $I$, $I=I_{\text{min}} \oplus \cdots \oplus I_{\text{max}}$. $I_{\text{max}}$ increases by $1/2$ at each step down in $Y$, so
that the states with $3Y = N_c$ have $I_{\text{max}}=(p+r)/2$.
Thus, 
the allowed isospin (and hence spin) states are $(p/2) \otimes (r/2)$.  {\it Case ($r \ge q$):} The states $r$ levels down from the upper edge are $s=p+q-r$ levels above the lower edge, so the solution is given by the previous case with $p \to q$, $r \to s \equiv p+q-r$.

In summary: The allowed $SU(3)$ Skyrme states are $(p,q)$ with
\begin{eqnarray}
p+2q=N_c+3r,\qquad
j = \left\{ \begin{array}{ll}{p \over 2} \otimes {r \over 2} & \text{if}\ r \le q, \\[5pt]
{q \over 2} \otimes {p+q-r \over 2} & \text{if}\  r \ge q .\\
\end{array}\right.
\label{32}
\end{eqnarray} 
Using the above equations modulo two shows that the soliton is a fermion or boson depending on whether $N_c$ is odd or even.

For $N_c=3$, the $r=0$ states have 3 boxes and are: $(1,1)  \to \mathbf{8}_{1/2}$ and $(3,0)  \to \mathbf{10}_{3/2}$. The $r=1$ states have 6 boxes and are $(0,3) \to 
\mathbf{\overline{10}}_{1/2}$, $(2,2) \to \mathbf{27}_{1/2},\mathbf{27}_{3/2}$, $(4,1) \to \mathbf{35}_{3/2},\mathbf{35}_{5/2}$, $(6,0) \to \mathbf{28}_{5/2}$, and similarly for higher values of $r$. The last case is an example where one needs to use the $r>q$ case. The $r \le q$ formula would give both $\mathbf{28}_{5/2}$ and $\mathbf{28}_{7/2}$ states.

For $F=2$, the only constraint is that $I=J$, so there is an infinite tower of states with all possible values $I=J=j$. Witten has argued that one must restrict to $2I=2J$ even states for $N_c$ even, and $2I=2J$ odd states for $N_c$ odd~\cite{Witten}.

The flavor Casimir is required to determine the rotational energy of the soliton.
The Casimir of the $SU(F)$ representation $w=(n_1,n_2,0\ldots,0,n_{-2},n_{-1})$ for $F \ge 5$ is
\begin{eqnarray}
C_2 &=& \frac12{n_1} \left(1 + \frac {n_1} 2 \right) + {N_c( N_c+2F)(F-2) \over 4 F} + 2 r^2 \label{11}  \\
&& \hspace{-1cm}+  \left( 2 F + N_c -4 \right) r + \left( \frac 52-2r-\frac12N_c-F \right)n_{-1} + \frac 3 4 n_{-1}^2 \nonumber
\end{eqnarray}
using the Freudenthal formula,
where Eq.~(\ref{6}) has been used to eliminate $n_{\pm 2}$ in favor of $N_c$ and $r$. The expression does not have a nice form, because $r$ is not symmetrically defined with respect to charge conjugation. Using the variables 
\begin{eqnarray}
j_q=n_1/2,\qquad j_{\bar q}=n_{-1}/2, \qquad E=n_{-1}+2n_{-2}
\label{15}
\end{eqnarray}
defined in Ref.~\cite{JM1}, instead of $n_{\pm1}$ and $r$, leads to the formula
\begin{eqnarray}
C_2 &=&  {N_c( N_c+2F)(F-2) \over 2 F} + E(E+2F+N_c-4) \nn
&& + j_q(j_q+1)   +  j_{\bar q}(j_{\bar q}+1)  \ .  
\label{12}
\end{eqnarray}

The variables $j_q$, $j_{\bar q}$ and $E$ arise naturally in a quark model construction. The baryon is made of $N_c+E$ quarks and $E$ antiquarks, where exoticness $E$ is the \emph{minimum} number of $q\bar q$ pairs required to construct a baryon with the desired flavor quantum numbers.  For $F\ge 5$, the quarks form the flavor representation $(n_1,n_2,0,\ldots,0)$ with $n_1+2n_2=N_c+E$, and spin $j_q=n_1/2$;  the antiquarks form the flavor representation $(0,\ldots,0,n_{-2},n_{-1})$ with $n_{-1}+2n_{-2}=E$ and spin $j_{\bar q}=n_{-1}/2$.  The exotic baryon obtained by combining the quarks and antiquarks has flavor weight $w= (n_1, n_2, 0, \ldots, 0, n_{-2}, n_{-1} )$, since flavor singlet $q \bar q$ pairs which can annihilate are excluded~\cite{JM1}. For $F\ge 5$, one can take the integers $n_{\pm 1}$, $n_{\pm 2}$ in the weight $w$ and construct $j_q, j_{\bar q}, E$. For $F < 5$, it is still useful to convert to quark model variables, even though the conversion is not as simple as Eq.~(\ref{15}), because the quark and antiquark contributions are not separated in the weight $w$.

For $F=4$, Eq.~(\ref{11}) continues to hold for $w=(n_1,n_0,n_{-1})$, where Eq.~(\ref{7}) has been used to eliminate $n_0$ in favor of $r$. One still can define $j_q=n_1/2$, $j_{\bar q}=n_{-1}/2$.  Exoticness $E= n_{-1} + 2 n_{-2}=2r-n_{-1}$ still can be determined from $w$ and $N_c$, and Eq.~(\ref{12}) remains valid.

For three flavors, the $(p,q)$ Casimir is
\begin{eqnarray}
C_2 &=& \frac 1 3 \left(p^2 + pq + q^2\right) + \left( p +  q \right) \ .
\label{17}
\end{eqnarray}
The weight $(p,q)$ and $N_c$ determine $p,q,r$. Eqs.~(\ref{pqr},\ref{15}) can be solved to give the $n_i$ in terms of $p,q,r,E$:
\begin{eqnarray}
&& n_1 = -E + p + r,\qquad n_{-1} = -E + 2 r  \nn
&& n_2 = E + q -2r,  \qquad n_{-2} =  E-r,
\end{eqnarray}
but now $E$ is not determined uniquely by $p,q$ and $N_c$. Since $n_i \ge 0$, one has
\begin{eqnarray}
 r \le E \le r+p , \qquad 2 r-q \le E \le 2r \ .
\label{16}
\end{eqnarray}
Definitions Eqs.~(\ref{pqr},\ref{15}) imply that $E=r+n_{-2}$, so that $E \not =r $~\cite{JM1,Kopeliovich}.
Since exoticness is the minimum value of $E$ for which one can construct a given baryon state in a quark model, Eq.~(\ref{16}) gives~\cite{JM1}
\begin{eqnarray}
E \! &=& \! E_{\text{min}} \!=\!  \max(r,2r-q) =   \left\{ \begin{array}{ll}r & \text{if}\ r \le q, \\
2r-q & \text{if}\ r \ge q .\\
\end{array}\right.
\end{eqnarray}
If $r \le q$, then $E=r$ gives $n_{-2}=0$, so $n_{-1}=E=r$, $n_1=p$, $n_2=q-r$,
and $j_q=p/2$ and $j_{\bar q}=r/2$. If $r \ge q$, then $E=2r-q$ gives $n_2=0$, so $n_1=p+q-r$, $n_{-1}=q$, $n_{-2}=r-q$, and
$j_q=(p+q-r)/2$ and $j_{\bar q}=q/2$. Combining $j_q$ and $j_{\bar q}$ gives the same spin states as Eq.~(\ref{32}).  Eq.~(\ref{12}) remains valid, and 
agrees with Eq.~(\ref{17}) for $r \le q$ and $ r \ge q$ (or equivalently, $n_{-2} \le n_2$ and $n_{-2} \ge n_2$).

For $F=2$, the quarks are in the representation $(2j)$ with spin $j$. In this case, for states with spin of order one, $E=r=0$, $j_q=j$, $j_{\bar q}=0$. The Casimir is
$C_2 = j(j+1)$, and Eq.~(\ref{12}) remains valid.

The rotational energy of the soliton is given by the Hamiltonian,
\begin{eqnarray}\label{18}
H = M_0 + {1 \over 2 I_1}\mathbf{J}^2 +  {1 \over 2 I_2} \left( 
\mathbf{T}^2 - \mathbf{J}^2 - {F-2 \over 4 F} N_c^2\right),
\end{eqnarray}
with corrections of order $1/N_c^2$. The mass $M_0$ and moments of inertia $I_{1,2}$ are order $N_c$. This collective coordinate quantization is valid for baryons with $E=0$, where the Casimir $\mathbf{T}^2$ is of order $N_c^0$, so that the rotational energy is order $1/N_c$. However, for $E \ne 0$ baryon exotics, the Casimir is order $N_c$, and the rotational energy is order $N_c^0$, which is the same order as the vibrational energies. In this case, one has to include vibrational-rotational mixing to correctly compute the energies, and Eq.~(\ref{18}) is no longer valid~\cite{IKOR}. Nevertheless, the quantum numbers of the soliton will not change, and the structure of the energy still has the form Eq.~(\ref{12}), though the coefficients of the $E$, $E^2$, $j_q(j_q+1)$ and $j_{\bar q}(j_{\bar q}+1)$ terms no longer have the values given in Eq.~(\ref{12})~\cite{JM1}.

This work was supported in part by the Department of Energy under Grant 
DE-FG03-97ER40546.

\end{document}